\documentclass[conference]{IEEEtran}
\ifCLASSINFOpdf
\else
\fi
\usepackage{balance}
\usepackage[square, comma, sort&compress, numbers]{natbib}
\usepackage{subcaption} 
\usepackage{graphicx}
\usepackage{caption}
\usepackage[table,xcdraw]{xcolor}
\usepackage{booktabs,graphicx}
\usepackage{multirow}

\begin{document}
\title{Arm Motion Classification Using Curve Matching of Maximum Instantaneous Doppler Frequency Signatures}

\author{Moeness G. Amin\\
Center for Advanced Communications\\
Villanova University\\
Villanova, PA 19085, USA \\
moeness.amin@villanova.edu
\and
Zhengxin Zeng, Tao Shan\\
School of Information and Electronics\\
Beijing Institute of Technology\\
Beijing, China \\
\{3120140293,shantao\}@bit.edu.cn \and
}


%Telephone: (800) 555--1212, Fax: (888) 555--1212}


\maketitle

\let\thefootnote\relax\footnotetext{The work of Mr. Zhengxin and Dr. Shan are funded by the International Graduate Exchange Program of Beijing Institute of Technology, and was performed while Mr. Zhengxin was a Visiting Scholar at the Center for Advanced Communications, Villanova University.}
\begin{abstract}
Hand and arm gesture recognition using the radio frequency (RF) sensing modality proves valuable in man-machine interface and smart environment. In this paper, we use curve matching techniques for measuring the similarity of the maximum instantaneous Doppler frequencies corresponding to different arm gestures. In particular, we apply both Fr\'echet and dynamic time warping (DTW) distances that, unlike the Euclidean (L2) and Manhattan (L1) distances, take into account both the location and the order of the points for rendering two curves similar or dissimilar. It is shown that improved arm gesture classification can be achieved by using the DTW method, in lieu of L2 and L1 distances, under the nearest neighbor (NN) classifier.
\end{abstract}
\renewcommand\IEEEkeywordsname{Keywords}

% \fi
\IEEEpeerreviewmaketitle

\section{Introduction}
Propelled by successes in discriminating between different human activities, radar has been recently employed for automatic hand gesture recognition for interactive intelligent devices \cite{li2018sparsity,kim2016hand,wang2016interacting,skaria2019hand,zhang2016dynamic,maminzz}. This recognition proves important in contactless close-range hand-held or arm-worn devices, such as cell phones and watches. The most recent project on hand gesture recognition, Soli, by Google for touchless interactions with radar embedded in a rest band is a testament of this emerging technology \cite{wang2016interacting}. In general, automatic hand or arm gesture recognition, through the use of radio frequency (RF) sensors, is important to smart environment. It is poised to make homes more user friendly and most efficient by identifying different motions for controlling instrument and household appliances. The same technology can greatly benefit the physically challenged who might be wheelchair confined or bed-ridden patients. The goal is then to enable these individuals to be self-supported and independently functioning.\par
Arm motions assume different kinematics than those of hands, especially in terms of speed and time duration. Compared to hand gesture, arm gesture recognition can be more suitable for contactless man-machine interaction with longer range separation, e.g. , the case of commanding appliances, like TV, from a distant couch. The larger radar cross-sections of the arms, vis-a-vis hands, permit more remote interactive positions in an indoor setting. Further, the ability of using hand gestures for device control can sometimes be limited by cognitive impairments such the Parkinson disease which induces strong hand tremor.\par
The arm is the part of the upper limb connected to the glenohumeral joint (shoulder joint) and the elbow joint. It can be divided into the upper arm, which extends from the shoulder to the elbow, the forearm, which extends from the elbow to the hand, and the hand. The nature and mechanism of arm motions are dictated by its elongated bone structure defined by the humerus which extends from the shoulder to the elbow and the radius and ulna that extend from elbow to hands. Because of such structure, arm motions, excluding hands, can be accurately simulated by two connected rods. In this respect, the instantaneous Doppler frequency of any point on the upper arm can be discerned from any other point in the same region. The same can be said for the forearm. This is different from hand motions which involve different motion of the palm and the fingers, and it is certainly different from body motions which yield intricate micro-Doppler (MD) signature \cite{amin2017radar,amin2016radar,seifert2019detection,van2008feature,kim2015human,mobasseri2009time,gurbuz2016micro,jokanovic2016radar,gurbuz2019radar}.\par
Recent work in automatic arm motion recognition using the maximum instantaneous Doppler frequencies, i.e., the envelope of the MD signature of the data spectrogram, as features followed by the NN classifier has provided classification rates reaching close to 97\% \cite{zeng2019automatic}. In this work, the feature vector consists of the augmented positive frequency and negative frequency envelopes. The corresponding classification performance outperformed data driven feature extraction, such as principal component analysis (PCA) and provided similar results to convolution neural networks. Since the NN classifier applies distance metrics to measure closeness of the test data to training data, only the envelope values rather than the actual shape of the envelope are used in classification.\par
In this paper, and towards improving on the results in \cite{maminzzfrance} , we employ features which capture the MD signature envelope behavior and its evolution characteristics. In particular, considering the envelope as a time-series or a curve, we measure the similarity between curves in a way that takes into account both the location and ordering of the points along the curve. It is noted that different measures for curve matching appear in several application domains, including time series analysis, shape matching, speech recognition, and signature verification. Curve matching has been studied extensively by computational geometry, and many measures of similarity have been examined \cite{Efrat2007}. We consider both the Fr\'echet distance and DTW distance which represent two of the most commonly curve matching metrics \cite{buchin2006computing,alt1995computing,munich1999continuous}. The Fr\'echest distance is a maximum measure over a parametrization, whereas the DTW is the sum-measure method. It is shown that the DTW-based NN classifier outperforms those based on L2 and L1 distance norms, and achieves an average classification rate above 99\%.  Similar to \cite{maminzzfrance}, our feature vector includes the augmented positive and negative frequency envelopes. But we also augment these two envelopes with a vector of their differences which captures the synchronization of the two envelopes.\par
The remainder of this paper is organized as follows. In Section II, we discuss the time-frequency signal representations and the power burst curve. Section III describes a method to extract the MD siganture envelopes. Section IV discusses the dynamic time warping method. Section V describes the radar data collection and pre-processing of arm motions. Section VI gives the classification results using real data measurements based on the DTW method. Section VII is the conclusion.

\section{Radar MD Signature Representation}
\subsection{Time-frequency Representations}
Arm motions generate non-stationary radar back-scattering signals, which are typically analyzed by time-frequency representations (TFRs). TFR reveals the signal local frequency behavior in the joint-variable domain referred to as the MD signature. A commonly used technique for TFRs is the spectrogram. For a discrete-time signal \(s(n)\) of length \(N\), the spectrogram can be obtained by taking the short-time Fourier transform (STFT) of the data and computing the magnitude square, 
\begin{equation}\label{stft}
              S\left( {n,k} \right) = {\left| {\sum\limits_{m = 0}^{L - 1} {s(n + m)h(m){e^{ - j2\pi \frac{{mk}}{N}}}} } \right|^2}\
\end{equation}
where \(n=1,\cdots,N\) is the time index, $k=1,\cdots\,K$ is the discrete frequency index, and $L$ is the length of the window function $h(\cdot)$. We deal with the MD signal as a deterministic signal rather than a stochastic process, and we do not assume any underlying frequency modulated signal model that calls for optimum parameter estimation \cite{Amin1992TimeFrequencySA}. Basically,  the spectrograms are used for feature extractions, and without considering any model for feature behaviors \cite{cirillo2008parameter}.
\subsection{Power Burst Curve (PBC)}
The onset and offset times of each motion can be determined by monitoring the PBC \cite{erol2017range,amin2019rf}, which measures the signal energy in the spectrogram within specific frequency bands. That is,
\begin{equation}\label{pbc}
S(n) = \sum\limits_{{k_1} = {K_{N1}}}^{{K_{N2}}} {{{\left| {S(n,{k_1})} \right|}^2}}  + \sum\limits_{{k_1} = {K_{P1}}}^{{K_{P2}}} {{{\left| {S(n,{k_1})} \right|}^2}} \
\end{equation}
The negative frequency indices $K_{N1}$ and $K_{N2}$ are set to $-500 Hz$ to $-20 Hz$, whereas the indices for positive frequencies are $K_{P1}=20 Hz$ and $K_{P2}=500 Hz$. The frequency band around the zero Doppler bin between $-20 Hz$ and $20 Hz$ affects the accuracy of the result and, therefore, is not considered.\par
A moving average filter is applied to smooth the original PBC curve. The filtered PBC is denoted as $S_{f}(n)$. The threshold, $T$, determines the beginning and the end of each motion, and is computed by
\begin{equation}\label{thre}
T = {S_{f\min }} + \alpha  \cdot \left( {{S_{f\max }} - {S_{f\min }}} \right)\
\end{equation}
where $\alpha $ depends on the noise floor and is empirically chosen from $[0.01,0.2]$. $S_{f\min}$ and $S_{f\max}$, respectively, represent the minimum and maximum values of $S_f(n)$. In our work, $\alpha $ is set to 0.1, which means 10\% over the minima. The onset time of each motion is determined as the time index at which the filtered PBC exceeds the threshold, whereas the offset time corresponds to the time index at which the filtered PBC falls below the threshold.\par
\section{Extraction of the MD Signature Envelopes}
We select features specific to the nominal arm motion local frequency behavior and power concentrations. These features are the positive and negative frequency envelopes in the spectrograms. The envelopes represent the maximum instantaneous Doppler frequencies. They attempt to capture, among other things, the maximum positive and negative Doppler frequencies, time-duration of the arm motion event and its bandwidth, the relative portion of the motion towards and away from the radar. In this respect, the envelopes can accurately characterize different arm motions. They can be determined by an energy-based thresholding algorithm discussed in \cite{maminzzfrance,erol2017range} and summarized below for convenience. First, the effective bandwidth of each motion is computed. This defines the maximum positive and negative Doppler frequencies. Second, the spectrogram is divided into positive frequency and negative frequency parts. The corresponding energies of the two parts, denoted as ${E_U}(n)$ and ${E_L}(n)$, are computed separately as,
 \begin{equation}\label{uenergy}
              {E_U}\left( n \right) = \sum\limits_{k = 1}^{\frac{K}{2} } {S{{\left( {n,k} \right)}^2}} ,  
               {E_L}\left( n \right) = \sum\limits_{k = \frac{K}{2}+1}^{K} {S{{\left( {n,k} \right)}^2}} 
\end{equation}
These energies are then scaled to define the respective thresholds, ${T_U}$ and ${T_L}$,
\begin{equation}\label{uthreshold}
              {T_U}(n) = {E_U}(n)\cdot {\sigma_U}, {T_L}(n) = {E_L}(n)\cdot {\sigma_L}
\end{equation}
where ${\sigma_U}$ and ${\sigma_L}$ represent the scale factors, both are less than 1. These scalars can be chosen empirically, but an effective way for their selections is to maintain the ratio of the energy to the threshold values constant over all time samples. This constant ratio can be found by time locating the maximum positive Doppler frequency and computing the corresponding energy at this location. Once the threshold is computed, the positive frequency envelope is then provided by locating the Doppler frequency at each time instant for which the spectrogram assumes the first higher or equal value to the threshold. This frequency, in essence, represents the effective maximum instantaneous Doppler frequency. Similar procedure can be followed for the negative frequency envelope.  The positive frequency envelope, ${e_U}(n)$, and negative frequency envelope, $e_L(n)$, are concatenated to form the feature vector $e=[{e_U},{e_L}]$. Since the alignment of the positive and negative envelopes is not captured in the concatenation of the envelope vectors, we also include their difference vector as a feature to define a new feature vector $e_{new}=[{e_U},{e_L},e_U-e_L]$.
\section{Dynamic Time Warping method}
The NN classifier is applied to the MD signature feature vector to discriminate among six arm motions. The performance is mainly determined by the distance metric used, where L1 and L2 norms are most common. On the other hand, the Fr\'echet distance and the DTW distance are two principal methods to calculate the similarity between two time series. The Fr\'echest distance is a max measure over a parametrization, whereas the DTW is the sum-measure method. In time series analysis, the DTW is an algorithm for measuring similarity between two temporal sequences which may vary in time or speed \cite{berndt1994using}. For instance, similarities in walking patterns could be detected using DTW, even if one person was walking faster than the other, or if there were accelerations and decelerations during the course of an observation.\par
Suppose $X=(x_1,x_2,\dots,x_i,\dots,x_n)$ and $Y=(y_1,y_2,\dots,y_j,\dots,y_n)$ are two time series representing the maximum instantaneous Doppler frequencies, an $n$-by-$n$ distance matrix $D$ is then formed, where the $(i,j)$ matrix element is the distance $D(x_i,y_j)$ between $x_i\in X$ and $y_j\in Y$ (the distance $D(x_i,y_j)$ is typically computed by the L1 or L2 norm). Each element also corresponds to an alignment between $x_i\in X$ and $y_j\in Y$. A warping path, $W$, finds a path in the distance matrix $D$,
\begin{equation}\label{path}
             W=w_1,w_2,\dots,w_l,\dots, w_L, n\le L\le 2n-1
\end{equation}
where each $w_l$ corresponds to an element $(i,j)_l$. The warping path is typically restricted by the following three constraints.

\textit{1) Boundary conditions:} The beginning and end of the path are $w_1=(1,1)$ and $w_L=(n,n)$ respectively.\par
\textit{2) Monotonicity:}  Given $w_{l1}=(a,b)$ and $w_{l2}=(c,d)$ where $a\le c$, we have $b\le d$.\par
\textit{3) Continuity:}  Given $w_{l}=(a,b)$ and $w_{l+1}=(c,d)$, we have $c-a\le 1 , d-b \le1$.\par
\begin{figure}[htbp]
% \vspace{-0.3cm}
\centering
\includegraphics[width=0.4\textwidth]{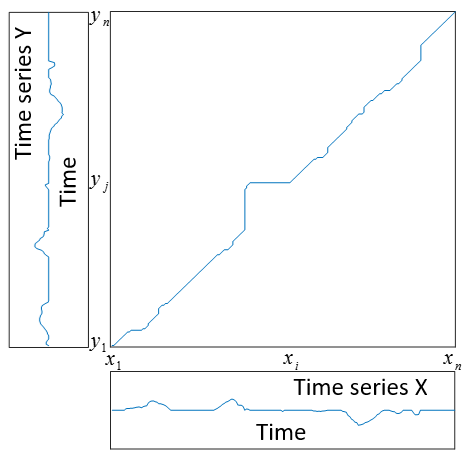}
\caption{An example of dynamic time warping path}
\label{warppingpath}
\end{figure}

The DTW is one of all possible paths that satisfy the above restrictions, and also has the minimum warping cost, as illustrated in Fig. \ref{warppingpath},
\begin{equation}\label{dtw}
            {{\rm{D}}_{{\rm{DTW}}}}(X,Y) = \min \sum\limits_{l = 1}^L {\left| {{w_l}} \right|} 
\end{equation}

\section{Arm Motion Experiments}
\begin{figure}[htbp]
% \vspace{-0.3cm}
\centering
\includegraphics[width=0.45\textwidth]{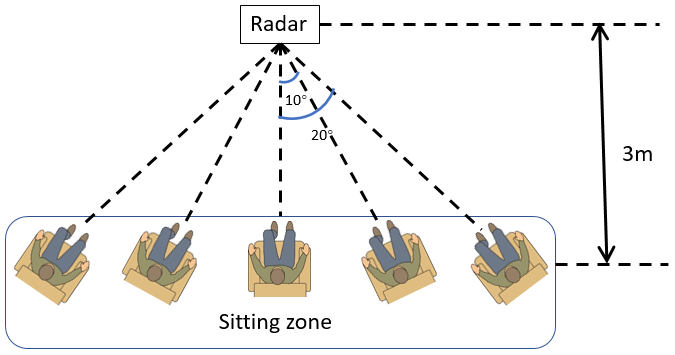}
\caption{Illustration of experiment setup}
\label{Sit}
% \vspace{-0.3cm}
\end{figure}

The data analyzed in this paper was collected in the Radar Imaging Lab at the Center for Advanced Communications, Villanova University. The radar system used in the experiments generates continuous wave, with carrier frequency and sampling rate equal to 25 GHz and 12.8 kHz, respectively. The radar was placed at the edge of a table. The arm motions were performed at approximately three meters away from radar in a sitting position of the participants. The body remained fixed as much as possible during the experiments. In order to mimic typical people behavior, the arms always rested down at a table or chair arm level at the initiation and conclusion of each arm motion. In the experiments, we chose five different orientation angles, $0,\pm 10^\circ,\pm 20^\circ$, as shown in Fig. \ref{Sit}, with the person always facing the radar. Different speeds of the arm motion are also considered.
\par

\begin{figure}[htbp]
% \vspace{-0.1cm}
% \setlength{\belowcaptionskip}{-0.1cm}
\begin{minipage}[b]{0.1\textwidth}
\includegraphics[width=1\textwidth]{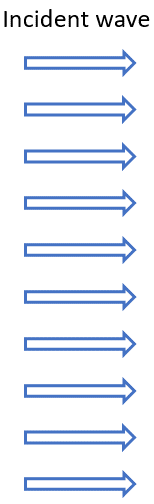} 
\vspace{0.1cm}
\end{minipage}
\begin{minipage}[b]{0.38\textwidth}
\includegraphics[width=1\textwidth]{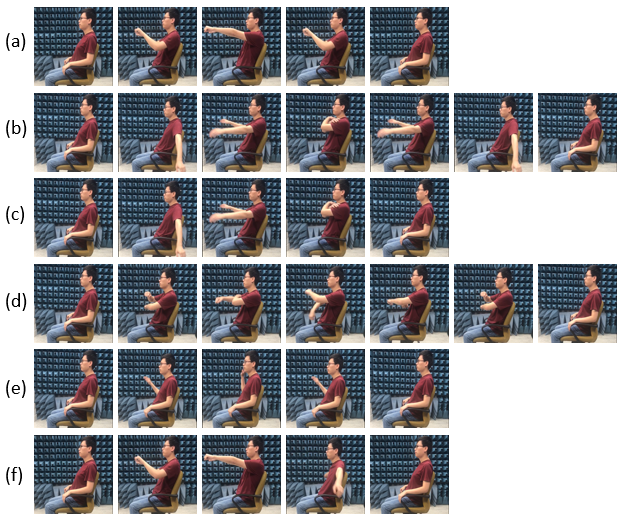}\\
%  \vspace{-0.5cm}
\end{minipage}
  \caption{ Illustrations of 6 different arm motions. (a) Pushing arms and pulling back, (b) Crossing arms and opening, (c) Crossing arms, (d) Rolling arms, (e) Stop sign, (f) Pushing arms and opening.} 
%   \figfooter{a}{Pushing arms and pulling back}
% \figfooter{b}{Crossing arms and opening}
% \figfooter{c}{Crossing arms}
% \figfooter{d}{Rolling arms}
% \figfooter{e}{Stop sign}
% \figfooter{f}{Pushing arms and opening}
\label{photo}
%  \vspace{-0.2cm}
\end{figure}

As depicted in Fig.\ref{photo}, the following six arm motions were conducted: (a) Pushing arms and pulling back, (b) Crossing arms and opening, (c) Crossing arms, (d) Rolling arms, (e) Stop sign, and (f) Pushing arms and opening. The description of these motions are as follows \cite{maminzzfrance}. In ``pushing," the arms moved towards the radar, whereas in ``pulling," they moved away from the radar. Both motions are relatively quick, with ``pulling" immediately following ``pushing." The motion of ``crossing arms" describes crossing the arms from a wide stretch. Six people participated in the experiment. Each arm motion was recorded over 40 seconds to generate one data segment. The recording was repeated 4 times, containing slow and normal motions at each angle. Each data segment contained 12 or 13 individual arm motions, and a 5 second time window is applied to capture the individual motions according to the onset and offset time determined by the PBC. In total, 1913 segments of data for six arm motions were generated. The most discriminative arm motion can be used as an ``attention'' motion for signaling the radar to begin, as well as to end, paying attention to the follow on arm motions. Among all arm motions, ``Pushing and open arms'' assumed the highest accuracy, and it was chosen as the ``attention'' motion.\par

\begin{figure}[htbp]
\begin{subfigure}[b]{0.5\linewidth} 
\centering\includegraphics[width=1\linewidth]{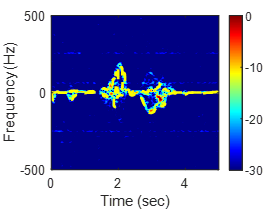} 
\captionsetup{justification=centering}
\caption{} 
\end{subfigure}\hfill
\begin{subfigure}[b]{0.5\linewidth} 
\centering\includegraphics[width=1\linewidth]{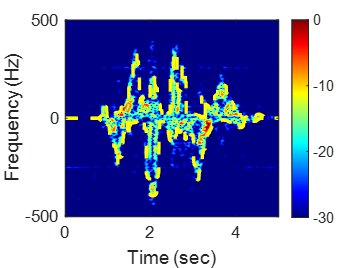} 
\captionsetup{justification=centering}
\caption{} 
\end{subfigure}\vspace{10pt}
\begin{subfigure}[b]{0.5\linewidth} 
\centering\includegraphics[width=1\linewidth]{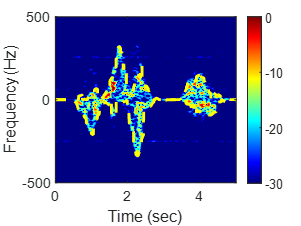} 
\captionsetup{justification=centering}
\caption{} 
\end{subfigure}\hfill
\begin{subfigure}[b]{0.5\linewidth} 
\centering\includegraphics[width=1\linewidth]{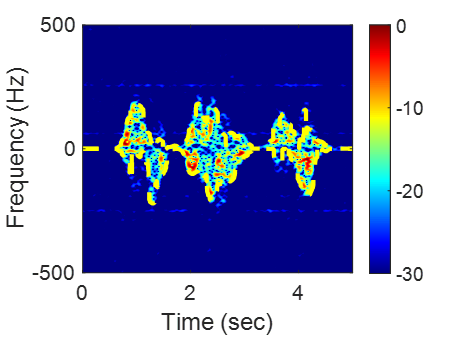} 
\captionsetup{justification=centering}
\caption{} 
\end{subfigure}\vspace{10pt}
\begin{subfigure}[b]{0.5\linewidth} 
\centering\includegraphics[width=1\linewidth]{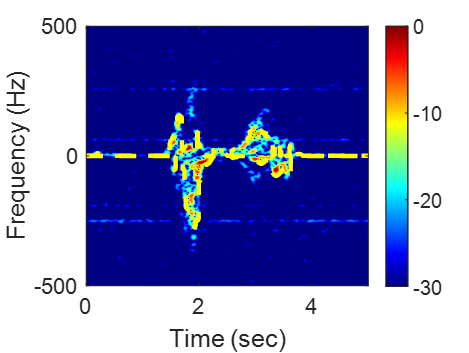} 
\captionsetup{justification=centering}
\caption{} 
\end{subfigure}\hfill
\begin{subfigure}[b]{0.5\linewidth} 
\centering\includegraphics[width=1\linewidth]{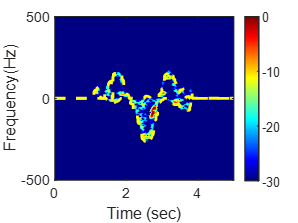} 
\captionsetup{justification=centering}
\caption{} 
\end{subfigure}
  \caption{ Spectrograms and corresponding envelopes. (a) Pushing arms and pulling back, (b) Crossing arms and opening, (c) Crossing arms, (d) Rolling arms, (e) Stop sign, (f) Pushing arms and opening.} 
%     \figfooter{a}{Pushing arms and pulling back}
% \figfooter{b}{Crossing arms and opening}
% \figfooter{c}{Crossing arms}
% \figfooter{d}{Rolling arms}
% \figfooter{e}{Stop sign}
% \figfooter{f}{Pushing arms and opening}
\label{envelopes}
%  \vspace{-0.2cm}
\end{figure}

Fig. \ref{envelopes} shows examples of spectrograms for six different arm motions with normal speed at zero angle. The employed sliding window $h(\cdot)$ is rectangular with length $L=$2048 (0.16 $s$), and $K$ is set to 4096. It is clear that the envelopes, representing the maximum instantaneous Doppler frequencies, can well enclose the local power distributions. It is also evident that the MD characteristics of the spectrograms are in agreement and consistent with each arm motion kinematics \cite{zeng2019automatic}. \par

\begin{figure}[htpb]
\centering\begin{subfigure}[b]{0.5\linewidth} 
\centering\includegraphics[width=1\linewidth]{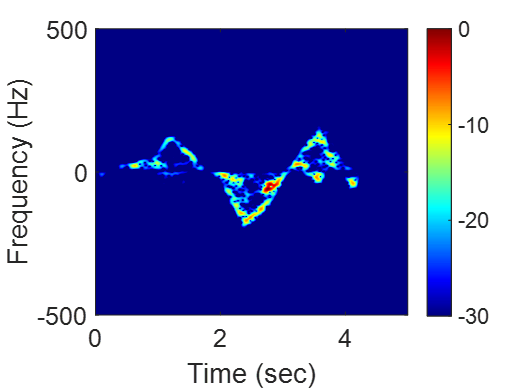} 
\captionsetup{justification=centering}
\caption{} 
\end{subfigure}\hfill
\begin{subfigure}[b]{0.5\linewidth} 
\centering\includegraphics[width=1\linewidth]{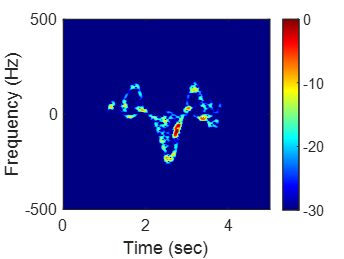} 
\captionsetup{justification=centering}
\caption{} 
\end{subfigure}
  \caption{ The ``attention'' motion with different velocities at $0^\circ$. (a) Slow motion, (b) normal motion.} 
% \leftline{\figfooter{a}{Slow motion}}
% \leftline{\figfooter{b}{normal motion}}
\label{velocity}
%  \vspace{-0.2cm}
\end{figure}

\begin{figure}[htpb]
\centering\begin{subfigure}[b]{0.5\linewidth} 
\centering\includegraphics[width=1\linewidth]{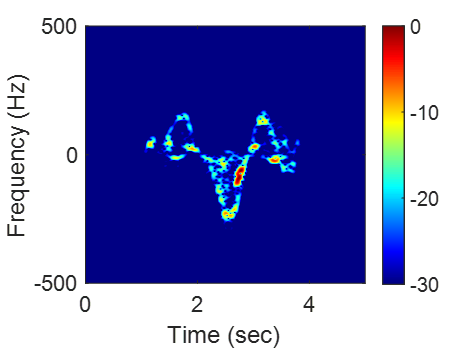} 
\captionsetup{justification=centering}
\caption{} 
\end{subfigure}\hfill
\begin{subfigure}[b]{0.5\linewidth} 
\centering\includegraphics[width=1\linewidth]{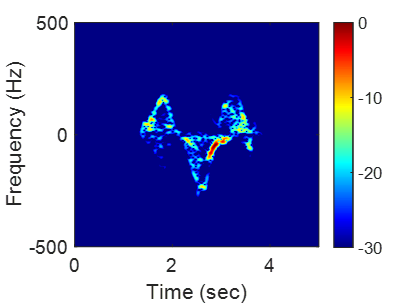} 
\captionsetup{justification=centering}
\caption{} 
\end{subfigure}\vspace{10pt}

\begin{subfigure}[b]{\linewidth} 
\centering\includegraphics[width=.5\linewidth]{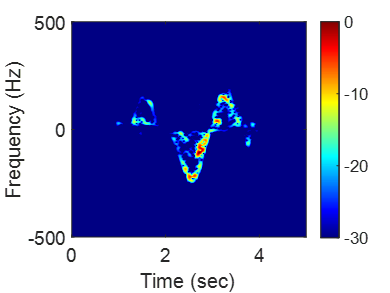} 
\captionsetup{justification=centering}
\caption{} 
\end{subfigure} 
  \caption{ The ``attention'' motion with normal speed at different orientation angles. (a) The ``attention'' motion at $0^\circ$, (b) The ``attention'' motion at $10^\circ$, (c) The ``attention'' motion at $20^\circ$.} 
% \leftline{\figfooter{a}{The ``attention'' motion at $0^\circ$}}
% \leftline{\figfooter{b}{The ``attention'' motion at $10^\circ$}}
% \leftline{\figfooter{c}{The ``attention'' motion at $20^\circ$}}
\label{angle}
\end{figure}

Fig. \ref{velocity} is an example of the ``attention'' motion with different velocities at $0^\circ$. The time period of the normal motion is shorter than that of the slow motion, and the speed is faster which causes higher Doppler frequencies. The main characteristics and behaviors, however, remain unchanged. Fig. \ref{angle} shows the ``attention'' motion with the normal speed at different orientation angles. As the angle increases, the energy becomes lower owing to the $dB$ drop in the antenna beam.

\section{Simulation}
In this section, All 1913 data segments are used to validate the proposed method where 70\% of the segments are used for training and 30\% for testing. The classification results are obtained by 500 Monte Carlo trials. \par
\begin{figure}[htpb]
\centering\begin{subfigure}[b]{1\linewidth} 
\centering\includegraphics[width=1\linewidth]{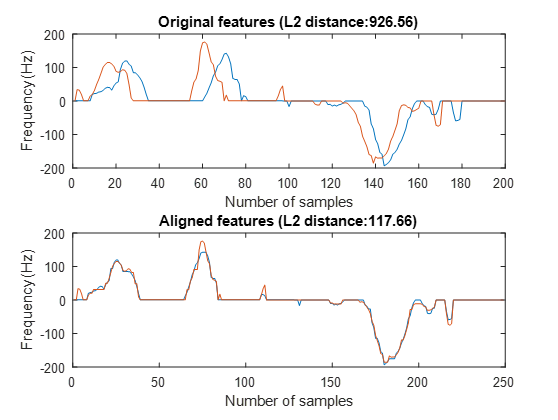} 
\captionsetup{justification=centering}
\caption{} 
\end{subfigure}\hfill
\begin{subfigure}[b]{1\linewidth} 
\centering\includegraphics[width=1\linewidth]{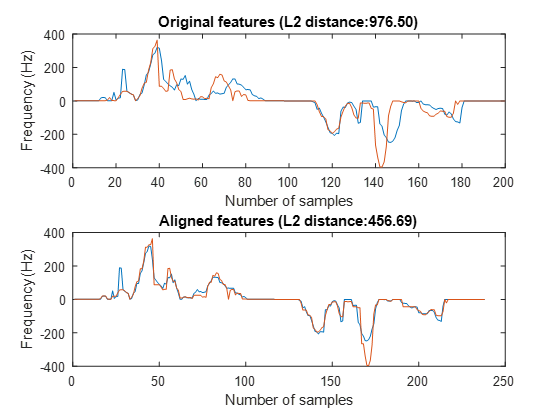} 
\captionsetup{justification=centering}
\caption{} 
\end{subfigure}
  \caption{ Alignment by the DTW. (a) Time shift, (b) Different speeds.} 
% \leftline{\figfooter{a}{Slow motion}}
% \leftline{\figfooter{b}{normal motion}}
\label{dtwalign}
%  \vspace{-0.2cm}
\end{figure}

The DTW distance accounts for the envelope misalignments in terms of time shift and time scaling (speed) of the data in the same motion class. In both cases, it yields high similarity measure, whereas the simple L1 and L2 distances reveal relatively lower similarity. Fig. \ref{dtwalign}(a) shows two envelopes of the same class with a time shift. It is clear that the DTW distance can align the two time series to reduce the effect of misalignments. Fig. \ref{dtwalign}(b) shows the alignment of the envelopes of two members of the same class but with different speeds. So, in this paper, we apply the DTW distance as the metric used by the NN classifier.\par
In our previous work \cite{zeng2019automatic,maminzzfrance}, each of the original extracted envelope feature contained 2000 samples for both positive and negative Doppler frequencies, and was directly fed into the NN classifier with L1 distance measure, achieving an overall accuracy 97.17\% \cite{zeng2019automatic}. To avoid high computations of DTW dealing with long time series, we downsample the envelopes to 200 samples. The downsampled envelope feature can maintain the main characteristics of the original envelope. To further examine the impact of downsampling on the NN classifier, the downsampled features are put into the NN classifier with L1 distance. This resulted in classification accuracy of 97.13\% \cite{zeng2019automatic}, which is nearly the same as when using the entire sequence. The confusion matrix is given in Table \ref{l1downcon}.\par

%Fig. \ref{downsample} shows one example of the envelope before and after downsampling. It is evident that t

%  \begin{figure}[htbp]
% \centering
% \includegraphics[width=0.4\textwidth]{downsample.png}
% \caption{An example of the envelope feature before and after downsampling.}
% \label{downsample}
% \end{figure}

\begin{table}[htbp]
\centering
\caption{ \sc Confusion Matrix Yielded by Envelope Method Based on NN-L1 Classifier}
\begin{tabular}{ccccccc}
\hline\hline
  & a       & b       & c       & d       & e       & f  \\ \hline
a &98.92\% &0 &0.02\%  & 0.01\%  & 1.04\%  & 0.01\%\\            
b &0.03\% & 95.28\% &2.62\% &0.03\% &0.45\%  & 1.59\% \\   
c &1.12\%& 0.24\%&95.74\% & 0.14\%  &2.28\%  &0.48\% \\ 
d &2.82\%  & 0 &0.59\% &95.78\% &0.81\% & 0\\
e &2.58\%  &0 &0.82\%  & 0  & 96.60\% & 0\\   
f &0.60\% &0.01\% & 0.05\%  &0 &0.56\% & 98.78\%\\ \hline \hline
\end{tabular}
\label{l1downcon}
\end{table}

\begin{table}[htbp]
\centering
\caption{ \sc Confusion Matrix Yielded by Envelope Method Based on NN-DTW Classifier}
\begin{tabular}{ccccccc}
\hline\hline
  & a       & b       & c       & d       & e       & f  \\ \hline
a &96.96\% &0 &0.02\%  & 0  & 2.79\%  & 0.23\%\\            
b &0.03\% & 98.70\% &0.71\% &0.07\% &0  & 0.45\% \\   
c &0.28\%& 0.38\%&97.87\% & 0  &1.39\%  &0.08\% \\ 
d &1.02\%  & 0 &1.42\% &96.82\% &0.59\% &0.15\%\\
e &0.17\%  &0 &0.48\%  & 0  & 99.09\% & 0.26\%\\   
f &0.13\% &0 & 0.04\%  &0 &0.69\% & 98.14\%\\ \hline \hline
\end{tabular}
\label{dtwcon}
\end{table}

With the downsampled envelope features, the NN classifier based on the DTW distance is applied. The result is an overall accuracy of 98.20\%, with the confusion matrix shown in Table \ref{dtwcon}. It took about 0.2 $s$ to classify each testing sample with the downsampled data using the DTW distance which is proper for real time processing. By comparing these two confusion matrices, the accuracy of motions (b), (c), (d) and (e) improves by 1\% to 3\%, whereas motion (a) dropped by 2\%. There is 1\% overall improvement. \par
The other curve matching method, Fr\'echest distance, is applied and achieves an average classification accuracy of only 85\%. The Fr\'echet distance only depends on the maximum length which leads to non-robust behavior, where small variations in the input can distort the distance function by a large amount. The DTW is the sum-measure rather than the max-measure and it takes into account small variations, which leads to a significantly better performance than the Fr\'echet distance.\par
As discussed in Section III, the new feature vector which includes the differences between the positive and negative envelopes is also analyzed by the DTW-based NN classifier. A remarkably higher average classification rate of 99.12\% is achieved. The confusion matrix shown in Table. \ref{dtwnewcon}. All motions are classified with an accuracy over 98.50\%, especially, the motion (c), (d) and (e) have an accuracy higher than 99\%.

\begin{table}[htbp]
\centering
\caption{ \sc Confusion Matrix Yielded by Envelope Method Based on NN-DTW Classifier with new feature vector}
\begin{tabular}{ccccccc}
\hline\hline
  & a       & b       & c       & d       & e       & f  \\ \hline
a &98.50\% &0 &0.01\%  & 0  & 1.45\%  & 0.04\%\\            
b &0.11\% & 98.80\% &0.55\% &0.01\% &0  & 0.53\% \\   
c &0.26\%& 0.29\%&99.10\% & 0  &0.31\%  &0.04\% \\ 
d &0.66\%  & 0 &0.15\% &99.15\% &0.03\% &0\\
e &0.01\%  &0 &1.01\%  & 0  & 98.97\% & 0.01\%\\   
f &0.02\% &0 & 0.01\%  &0 &0.32\% & 99.65\%\\ \hline \hline
\end{tabular}
\label{dtwnewcon}
\end{table}

\section{Conclusion}
We introduced a simple and practical technique for effective automatic arm motion recognition based on radar MD signature envelopes. No range or angle information was incorporated in the classifications. An energy-based thresholding algorithm was applied to separately extract the positive and negative frequency envelopes of the signal spectrogram. The feature vector is the augmented positive and negative frequency envelopes, and their difference vector. The augmented feature vector was provided to the NN classifier, and the DTW distance, which is more suitable to describe the similarity between curves, was employed in lieu of the L1 and L2 distance measures. It was shown that the NN classifier based on the DTW distance achieves close to 99 percent classification rate which is superior to existing work based on L1 distance by an overall 2\% improvement.





%\bibliographystyle{IEEEtran}


% \end{thebibliography}

\footnotesize
\balance
\bibliographystyle{IEEEtran.bst}
\bibliography{IEEEabrv,refs.bib}

\end{document}